\documentclass[aps,pra,preprint,groupedaddress,amsmath,amssymb,showpacs]{revtex4-1}
\usepackage{graphicx,mathrsfs,epsfig,amsmath,amssymb,multirow}
\usepackage{dcolumn}
\usepackage{bm}
\allowdisplaybreaks
\def\be{\begin{equation}}
\def\ee{\end{equation}}
\def\bea{\begin{eqnarray}}
\def\eea{\end{eqnarray}}
\def\bes{\begin{subequations}}
\def\ees{\end{subequations}}
\usepackage[usenames]{color}

\begin{document}
\title{Trapping of Weak Signal Pulses by Soliton and Trajectory Control in a Coherent Atomic Gas}
\author{Zhiming Chen}
\affiliation{State Key Laboratory of Precision Spectroscopy and Department of Physics,
East China Normal University, Shanghai 200062, China}

\author{Guoxiang Huang}
\email[Email: ]{gxhuang@phy.ecnu.edu.cn}
\affiliation{State Key Laboratory of Precision Spectroscopy and Department of Physics,
East China Normal University, Shanghai 200062, China}
\date{\today}

\begin{abstract}

We propose a method for trapping weak signal pulses by soliton and realizing its trajectory control via electromagnetically induced transparency (EIT). The system we consider is a cold, coherent atomic gas with a tripod or multipod level configuration.
We show that, due to the giant enhancement of Kerr nonlinearity contributed by EIT,
several weak signal pulses can be effectively trapped by a soliton  and cotravel stably with ultraslow propagating velocity. Furthermore, we demonstrate that the trajectories
of the soliton and the trapped signal pulses can be manipulated by using a Stern-Gerlach gradient magnetic field. As a result, the soliton and the trapped signal pulses
display a Stern-Gerlach deflection and both of them can bypass an obstacle together.
The results predicted here may be used to design all-optical switching at very low light level.

\end{abstract}


\maketitle

\section{Introduction}{\label{Sec:1}}

In recent years, the technique of trapping material (or massive) particles by light has been successfully developed and widely used in many research fields~\cite{met,dho,woe}. Because there is no interaction between photons in vacuum, for trapping light by light one must resort to optical media. The principle of trapping light by light is that a light beam acting on an optical medium induces a change of refractive index, which may provide an attractive potential to trap another light beam~\cite{Agrawal}.

Recently, much attention has been paid to the study on soliton-radiation trapping (SRT), in which a localized nonlinear optical beam, i.e., optical soliton, traps a weak optical beam through cross-phase modulation (CPM) effect. SRT can also occur for optical pulses if the condition of group-velocity matching between soliton and weak pulse is fulfilled~\cite{Hasegawa,NG,GS1,GS2}.

The most typical example of the SRT occurs in optical fibers. The recent development of highly nonlinear fibers has led to the observation of many nonlinear optical effects such as the Raman-induced frequency shifts and supercontinuum generation~\cite{Agrawal,DGC}. These nonlinear phenomena have many new applications in fields as diverse as high-precision metrology and optical coherence tomography. When an optical soliton is launched in a nonlinear fiber, it may emit a dispersive wave, called nonsolitonic or Cherenkov radiation, with propagation constant (or phase velocity) matched with that of the soliton. If the group velocities of both the soliton and the nonsolitonic radiation are close, the phenomenon of SRT appears, which has promising applications for realizing, e.g., different types of optical switching and supercontinuum generation~\cite{Agrawal,NG,GS1,GS2,DGC}.

There is much research on SRT in various physical settings~\cite{NG,GS1,GS2,DGC,gor1,gor2,liu,san,bat}. In a recent work, Saleh and Biancalana~\cite{MF} proposed a technique for obtaining an optical pulse trapping (similar to SRT) in which a soliton traps a small-amplitude optical pulse in a symmetric hollow-core photonic crystal fiber filled with a noble gas. However, up to now all works on optical pulse trapping mentioned above utilized passive optical media, in which far-off resonance excitation schemes are employed for avoiding significant optical absorption. To obtain optical pulse trapping in passive media, very high light-intensity and ultrashort laser pulses are required to obtain nonlinearity strong enough to balance the dispersion and/or diffraction effects; furthermore, an active control on the property of optical pulse trapping is not easy to realize in passive media because of the absence of energy-level structure and selection rules that can be used and manipulated.

In this article, we propose a mechanism for trapping weak signal pulses by soliton and realize its trajectory control via electromagnetically induced transparency (EIT), a typical quantum interference effect occurring in resonant multilevel systems~\cite{fle}. The system we consider is a cold, coherent atomic gas with a tripod or multipod level configuration.
By means of EIT, not only the optical absorption can be largely suppressed but also an enhanced Kerr nonlinearity can be obtained. We show that several weak signal pulses can be easily trapped by a soliton and stably cotravel with an ultraslow  propagating velocity. Furthermore, we demonstrate that the trajectories of the soliton and the trapped signal pulses can be steered by using a Stern-Gerlach (SG) gradient magnetic field, a technique used recently for linear
optical beams~\cite{Weitz,Sun1,Sun2} and light bullets~\cite{Huang1,Huang2}. As a result, the soliton and the trapped signal pulses display a SG deflection and both of them can bypass an obstacle together. The results predicted here may have potential applications for optical information processing, such as for the design of all-optical switching at very low light level.

The article is arranged as follows. In Sec.~\ref{Sec:2}, the theoretical model under study is described. In Sec.~\ref{Sec:3}, the nonlinear envelope equations governing the evolution of the probe and signal pulses are derived. The trapping of weak signal pulses by soliton and its SG deflection are investigated in detail in Sec.~\ref{Sec:4} and Sec.~\ref{Sec:5}, respectively. Finally, in Sec.~\ref{Sec:6} a summary of our main results is given.


\section{Model}{\label{Sec:2}}

\begin{figure}
\includegraphics[scale=0.6]{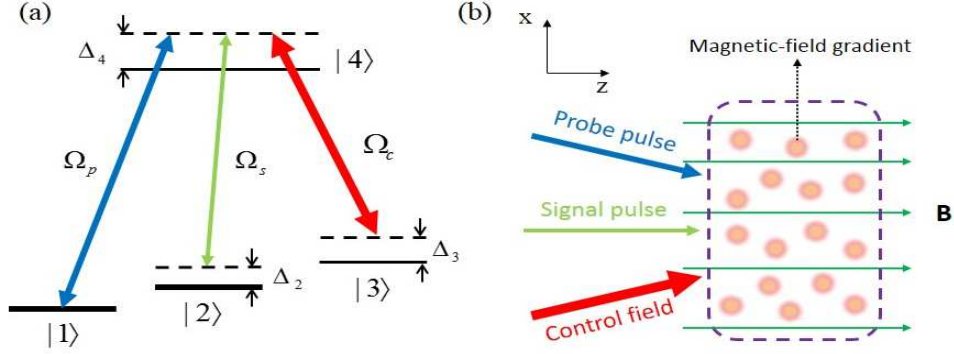}
\caption{(Color online) (a) Tripod-type atomic level diagram and excitation scheme.
$\Delta_2$ and $\Delta_3$ are two-photon detunings and $\Delta_4$ is one-photon detuning.
$\Omega_p$, $\Omega_s$, and $\Omega_c$ are half Rabi frequencies of the probe, signal, and control fields, respectively. (b) Possible experimental arrangement of beam geometry.
The probe, signal, and control fields propagate nearly along the $z$ direction~\cite{note}. A Stern-Gerlach magnetic field is applied along the $z$ direction with its gradient along the $x$ direction. The region in the closed dashed line indicates the ultracold atomic gas.}\label{fig:1}
\end{figure}

We consider a lifetime-broadened atomic gas with a tripod-type level configuration,  interacting resonantly with three laser fields, i.e., pulsed probe (with half Rabi frequency $\Omega_p$), pulsed signal (with half
Rabi frequency $\Omega_s$), and continuous-wave control (with half Rabi frequency $\Omega_c$)  fields.
The probe field has center frequency $\omega_{p}/(2\pi)$  and couples with the
$|1\rangle \leftrightarrow |4\rangle$ transition, the signal field has the center frequency
$\omega_{s}/(2\pi)$  and couples with the $|2\rangle \leftrightarrow |4\rangle$
transition, and the control field has the center frequency $\omega_{c}/(2\pi)$
and couples with the $|3\rangle \leftrightarrow |4\rangle$ transition, respectively
[Fig.~\ref{fig:1}(a)\,].
%
%
We assume atoms are cooled to an ultralow temperature so that their center-of-mass motion is negligible.

For simplicity, we assume that all the laser fields propagate nearly along $z$ direction~\cite{note}.
Thus the electric field vector in the system
reads $\textbf{E}=\sum_{l=p,s,c}\textbf{e}_l{\cal E}_l e^{i(k_l z-\omega_l t)}+\textrm{c.c.}$, where $\textbf{e}_l$
(${\cal E}_l$) is the unit polarization vector (envelope) of the $l$th polarization component. Additionally,
we assume a SG gradient magnetic field,
\begin{eqnarray}\label{B}
\textbf{B}(x,t)={\bf e}_z B x,
\end{eqnarray}
is applied to the system, where ${\bf e}_z$ is the unit vector in the
$z$-direction and $B$ characterizes the transverse gradient.
Due to the presence of the SG gradient magnetic field, the Zeeman level shift $\Delta
E_{j,\textrm{Zeeman}}=\mu_Bg_F^{j}m_F^{j}Bx$ for the level $E_j$ occurs. Here $\mu_B$, $g_F^{j}$, and $m_F^{j}$ are Bohr magneton, gyromagnetic factor, and magnetic quantum number of the level $|j\rangle$, respectively.
The propose of introducing the SG gradient magnetic field  is to produce an external force in transverse (i.e. $x$, $y$) directions so that the traveling trajectory of the probe and signal optical pulses can be manipulated, as will be shown below.

Under electric-dipole and rotating-wave approximations, the Hamiltonian of the system in the interaction picture reads
\bea
\hat{\mathcal{H}}_{\textrm{int}}= & &
-\sum_{j=1}^{4}\hbar\Delta_j|j\rangle\langle
j|-\hbar\left[\Omega_{p}|4\rangle\langle
1|+\Omega_{s}|4\rangle\langle
2|+\Omega_{c}|4\rangle\langle 3|+\textrm{H.c.}\right],
\eea
with $\Delta_1=0$, $\Omega_{p}=({\textbf e}_{p}\cdot{\textbf p}_{14}){\cal E}_{p}/\hbar$, $\Omega_{s}=({\textbf e}_{s}\cdot{\textbf p}_{24}){\cal E}_{s}/\hbar$, and $\Omega_{c}=({\textbf e}_c\cdot {\textbf p}_{34}){\cal E}_c/\hbar$. Here ${\textbf p}_{jl}$ is the electric-dipole matrix element related to the states $|j\rangle$ and $|l\rangle$.

The equation of motion for density matrix $\sigma$ in the interaction picture is~\cite{boyd}
\begin{equation}\label{DME}
\frac{\partial \sigma}{\partial t}=-\frac{i}{\hbar}[\hat{\mathcal{H}}_{\rm
int},\sigma]-\Gamma \sigma,
\end{equation}
where $\Gamma$ is a relaxation matrix. The explicit expression for density-matrix elements $\sigma_{jl}$ is given in Appendix~{\ref{AppendixA}.

The equation of motion for probe-field and signal-field Rabi frequencies $\Omega_p$ and $\Omega_s$ can be captured by the Maxwell equation $\nabla^2 \textbf{E}-(1/c^2)\partial^2 \textbf{E}/\partial t^2=[1/(\epsilon_0c^2)]\partial^2 \textbf{P}/\partial t^2$, where $\textbf{P}={\cal N}_a\textrm{Tr}(\textbf{p}\rho)$ with $\mathcal{N}_a$ the atomic
concentration and $\rho$ the density matrix in the Schr\"{o}dinger picture. Under slowly varying envelope approximation,
the equations of motion for $\Omega_p$ and $\Omega_s$ read~\cite{HDP}
\begin{subequations}\label{ME}
\begin{eqnarray}
&&i\left(\frac{\partial}{\partial
z}+\frac{1}{c}\frac{\partial}{\partial
t}\right)\Omega_{p}+\frac{c}{2\omega_{p}}\frac{\partial
^2}{\partial x^2}\Omega_{p}+\kappa_{14}\sigma_{41}=0,\\
&&i\left(\frac{\partial}{\partial
z}+\frac{1}{c}\frac{\partial}{\partial
t}\right)\Omega_{s}+\frac{c}{2\omega_{s}}\frac{\partial
^2}{\partial x^2}\Omega_{s}+\kappa_{24}\sigma_{42}=0,
\end{eqnarray}
\end{subequations}
where $\kappa_{14}={\cal N}_a\omega_{p}|\textbf p_{14}\cdot \textbf e_{p}|^2/(2\epsilon_0c\hbar)$ and $\kappa_{24}={\cal N}_a\omega_{s}|\textbf p_{24}\cdot \textbf e_{s}|^2/(2\epsilon_0c\hbar)$, with $c$ the light speed in vacuum. Note that we have assumed $R_x\ll R_y$, where $R_x$ and $R_y$
are respectively the transverse radii of the probe and signal pulses, and hence
the diffraction in the $y$ direction (i.e., $\partial \Omega_{p,s}/\partial y^2$) is negligible.
Additionally, we have assumed the control field is strong enough
so that $\Omega_c$  can be regarded as a constant during the evolution of the probe and signal pulses.

The model given above can be easily realized by selecting realistic physical systems. One of them is the ultracold $^{87}$Rb atomic gas with the energy levels selected as $|1\rangle=|5^{2}S_{1/2}, F=1, m_F=-1\rangle$ ($g_F=-1/2$), $|2\rangle=|5^{2}S_{1/2}, F=1, m_F=0\rangle$ ($g_F=-1/2$), $|3\rangle=|5^{2}S_{1/2}, F=2, m_F=0\rangle$ ($g_F=1/2$), and $|4\rangle=|5^{2}P_{1/2}, F=2, m_F=0\rangle$ ($g_F=1/6$)~\cite{Steck}. The decay rates are given by $\Gamma_2\simeq\Gamma_3\simeq1$ kHz and $\Gamma_4=5.75$ MHz. If the atomic concentration $\mathcal{N}_a$ is taken to be $3.67\times10^{10}$ cm$^{-3}$, we have $\kappa_{14}\approx\kappa_{24}\approx1.0\times10^9$ cm$^{-1}$s$^{-1}$.
These system parameters will be used in the following calculations.


\section{Nonlinear envelope equations and giant Kerr effect} {\label{Sec:3}}

We first use the standard method of multiple scales~\cite{HDP} to derive nonlinear envelope equations of the probe and signal pulses based on the Maxwell-Bloch Eqs.~(\ref{DME}) and~(\ref{ME}). To this end, we take the asymptotic expansion $\sigma_{jl}=\sigma_{jl}^{(0)}+\epsilon\sigma_{jl}^{(1)}+\epsilon^{2}\sigma_{jl}^{(2)}+\cdots$,
$d_{jl}=d_{jl}^{(0)}+\epsilon d_{jl}^{(1)}+\epsilon^{2}d_{jl}^{(2)}$ ($j,\,l=1,2,3,4$),
$\Omega_{p}=\epsilon \Omega_{p}^{(1)}+\epsilon^{2}\Omega_{p}^{(2)}+\epsilon^{3}\Omega_{p}^{(3)}+\cdots$, and
$\Omega_{s}=\epsilon^{2}\Omega_{s}^{(2)}+\epsilon^{3}\Omega_{s}^{(3)}+\cdots$.
Here $\sigma _{jj}^{(0)}$ is the initial population distribution prepared in the state $|j\rangle$, which is assumed as $1/2$ ($j=1,\,2$) for simplicity; $\epsilon$ is a dimensionless small parameter characterizing the typical amplitude of the probe pulse. All the quantities on the right-hand side of the expansion are considered as functions of the multiscale variables $x_1=\epsilon x$, $z_{\alpha}=\epsilon^{\alpha}z$ ($\alpha=0,\,2,\,3$), and $t_{\alpha}=\epsilon^{\alpha}t$ ($\alpha=0,\,2$). Additionally, the SG gradient magnetic field~(\ref{B}) is assumed to be of the order of $\epsilon^{2}$. Thus we have $d_{jl}^{(0)}=\delta_j-\delta_l+i\gamma_{jl}$, $d_{jl}^{(1)}=0$, and $d_{jl}^{(2)}=-\mu_{jl}Bx_1$. Note that because we are looking for
trapping of weak signal pulse by a soliton, i.e., the order of magnitude of  $\Omega_s$ is lower than that of  $\Omega_p$,
the expansion of $\Omega_p$ ($\Omega_s$) is assumed to start from the order of  $\epsilon$  ($\epsilon^2$).

Substituting the expansion into Maxwell-Bloch Eqs.~(\ref{DME}), and (\ref{ME}) and comparing the coefficients of $\epsilon^\alpha$ ($\alpha=1,\,2,\,3,\,\ldots$), we obtain a set of linear but inhomogeneous equations which can be solved order by order.
At the first order ($\alpha=1$), we obtain $\Omega_{p}^{(1)}=F_{1}e^{i\theta_p}$
with $\theta_p=K_{p}(\omega)z_0-\omega t_0$~\cite{note1} and $F_{1}$, a yet to be determined envelope function.
The linear dispersion relation $K_p(\omega)$ is given by
\begin{eqnarray}\label{Kp}
K_p(\omega)=\frac{\omega}{c}+\kappa_{14}\frac{(\omega+d_{31}^{(0)})\sigma_{11}^{(0)}}{|\Omega_c|^2-(\omega+d_{31}^{(0)})(\omega+d_{41}^{(0)})}.
\end{eqnarray}

Shown in Fig.~\ref{fig:2}(a) and Fig.~\ref{fig:2}(b) are, respectively, the imaginary part Im($K_p$) and real part Re($K_p$) of $K_p$  as functions of $\omega$. The dashed and solid lines in the figure correspond to the presence ($\Omega_c=1.0\times10^7$ s$^{-1}$) and the absence ($\Omega_c=0$) of the control field, respectively. We see that when $\Omega_c=0$ the probe pulse suffers a large absorption around $\omega=0$ [the solid line in Fig.~\ref{fig:2}(a)\,]. Nevertheless, when $\Omega_c\neq 0$  a transparency window is opened in Im($K_p$) [the dashed line in Fig.~\ref{fig:2}(a)\,], and hence the probe pulse can propagate in the resonant atomic system with negligible absorption, a basic character of EIT. On the other hand, when the EIT occurs the slope of Re($K_p$) is drastically changed and steepened [see the dashed line in Fig.~\ref{fig:2}(b)\,], which results in a significant reduction of the group velocity of the probe and signal pulses. All these important characters are due to the quantum interference effect induced by the control field.
%
\begin{figure}
\includegraphics[scale=0.7]{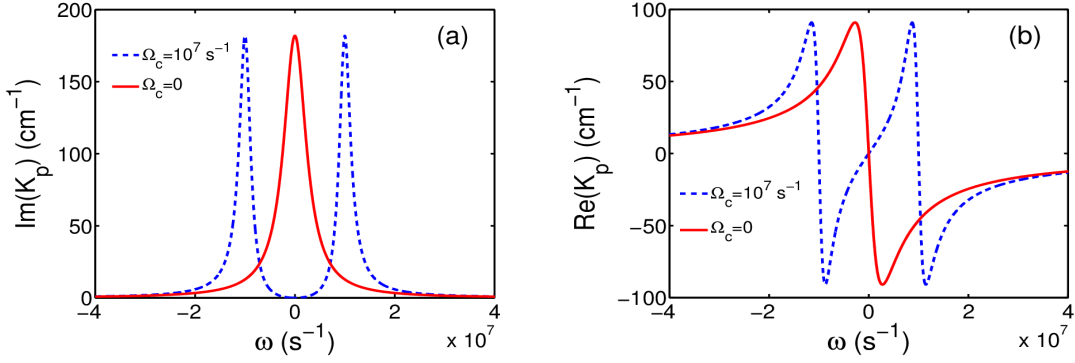}
\caption{(Color online) Linear dispersion relation of the probe field. (a) Im($K_p$) and (b) Re($K_p$) as functions of $\omega$. The dashed and solid lines  correspond to the presence ($\Omega_c=1.0\times10^7$ s$^{-1}$) and the absence ($\Omega_c=0$) of the control field, respectively.}\label{fig:2}
\end{figure}
%

At the second order ($\alpha=2$) we obtain $\Omega_{s}^{(2)}=F_{2}e^{i\theta_s}$,
where $\theta_s=K_{s}(\omega)z_0-\omega t_0$ with $F_{2}$ the envelope function yet to be determined and
\begin{eqnarray}\label{Ks}
K_s(\omega)=\frac{\omega}{c}+\kappa_{24}\frac{(\omega+d_{32}^{(0)})\sigma_{22}^{(0)}}{|\Omega_c|^2-(\omega
+d_{32}^{(0)})(\omega+d_{42}^{(0)})},
\end{eqnarray}
which is the linear dispersion relation of the signal pulse. Note that the property of $K_s(\omega)$
is similar to $K_p(\omega)$, i.e. it has an EIT transparency window in the imaginary part Im($K_s$)
for $\Omega_c\neq 0$ .

At the third order ($\alpha=3$), a solvability condition yields the equation for $F_1$:
\begin{eqnarray}\label{NLSE1}
i\left(\frac{\partial}{\partial
z_2}+\frac{1}{V_{\textrm{g}1}}\frac{\partial}{\partial
t_2}\right)F_{1}+\frac{c}{2\omega_{p}}\frac{\partial ^2}{\partial
x_1^2}F_{1}-W_{11}|F_{1}|^2F_{1}e^{-2\bar{a}_1z_2}+M_1 B x_1F_1=0,
\end{eqnarray}
where $\bar{a}_1=\epsilon^{-2}{\rm Im}(K_{p0})$, $W_{11}$
is a nonlinear coefficient related to the Kerr effect describing the
self-phase modulation of the probe pulse,
and the term related to $M_1$ denotes the contribution by the SG gradient magnetic field.

At the fourth order ($\alpha=4$), we obtain the equation for $F_2$:
\begin{eqnarray}\label{NLSE2}
i\left(\frac{\partial}{\partial
z_2}+\frac{1}{V_{\textrm{g}2}}\frac{\partial}{\partial
t_2}\right)F_{2}+\frac{c}{2\omega_{s}}\frac{\partial ^2}{\partial
x_1^2}F_{2}-W_{21}|F_{1}|^2F_{2}e^{-2\bar{a}_1z_2}+M_2 Bx_1F_2=0,
\end{eqnarray}
where $W_{21}$ is the nonlinear coefficient related to the cross-phase
modulation (CPM) contributed by the probe field and the term related to $M_2$
comes from the SG gradient magnetic field.

Equations (\ref{NLSE1}) and (\ref{NLSE2}) can be written as
\begin{eqnarray}\label{NLSE}
&&i\left(\frac{\partial}{\partial
z}+\frac{1}{V_{\textrm{g}p}}\frac{\partial}{\partial
t}\right)U_{1}+\frac{c}{2\omega_{p}}\frac{\partial ^2 U_{1}}{\partial
x^2}-W_{11}|U_{1}|^2U_{1}+M_1 BxU_1=0\\
&&i\left(\frac{\partial}{\partial
z}+\frac{1}{V_{\textrm{g}s}}\frac{\partial}{\partial
t}\right)U_{2}+\frac{c}{2\omega_{s}}\frac{\partial ^2 U_{2}}{\partial
x^2}-W_{21}|U_{1}|^2U_{2}+M_2 BxU_2=0
\end{eqnarray}
($j=1,\,2$).  The detailed expressions
of $W_{11}$, $W_{21}$, $M_1$, $M_2$, and each-order approximation solutions in the asymptotic
are presented in Appendix~\ref{AppendixB}.

We now estimate the self- and cross-Kerr susceptibilities based on the formulas of
$W_{11}$ and $W_{21}$ given by Eqs.~(\ref{W11}) and  (\ref{W21}). We obtain
\begin{subequations}
\begin{eqnarray}
&&\chi_{pp}^{(3)}=\frac{2c}{\omega_p}\frac{|\textbf{p}_{14}|^2}{\hbar^2} W_{11},\\
&&\chi_{sp}^{(3)}=\frac{2c}{\omega_s}\frac{|\textbf{p}_{14}|^2}{\hbar^2} W_{21}.
\end{eqnarray}
\end{subequations}
Using the system parameters given at the end of the last section and choosing
$\Omega_c=3.0\times10^7$ s$^{-1}$, $\delta_2=1.0\times10^4$ s$^{-1}$, $\delta_3=1.0\times10^5$ s$^{-1}$, and $\delta_4=1.6\times10^7$ s$^{-1}$, we obtain $\chi_{pp}^{(3)}\approx(4.95+0.89i)\times10^{-4}$ cm$^2$V$^{-2}$ and  $\chi_{sp}^{(3)}\approx(4.94+0.89i)\times10^{-4}$ cm$^2$V$^{-2}$.
We see that there are two features for  $\chi_{pp}^{(3)}$ and  $\chi_{ps}^{(3)}$. First, their imaginary parts are
much less than the real parts, contributed by the EIT effect. Second, their real parts have the order of magnitude
$10^{-4}$ cm$^2$V$^{-2}$, which is $10^{11}$ times larger than the third-order susceptibilities for conventional nonlinear optical media~\cite{Saltiel}. The physical reason for the enhancement of the Kerr nonlinearity in the present system is originated from the fact that the system is highly resonant and works under the
EIT condition.

For convenience, we convert Eq.~(\ref{NLSE}) into the dimensionless form
\begin{eqnarray}\label{NLSE3}
&&i\left(\frac{\partial}{\partial
s}+\frac{1}{\lambda_j}\frac{\partial}{\partial
\tau}\right)u_{j}+\frac{1}{2}\frac{\partial ^2}{\partial
\xi^2}u_{j}-w_{j1}|u_{1}|^2u_{j}+\mathcal{M}_j \xi u_j=-iA_{j}u_{j}
\end{eqnarray}
($j=1,\,2$), with $u_{1(2)}=(\Omega_{p(s)}/U_0)\exp[-i\textrm{Re}(K_{p(s)0})z]$, $s=z/L_{\textrm{Diff}}$, $\tau=t/\tau_0$, $\xi=x/R_x$, $\lambda_j=V_{\textrm{g}j}\tau_0/L_{\textrm{Diff}}$, $w_{j1}=W_{j1}/|W_{11}|$, $\mathcal{M}_j=M_jL_{\rm Diff}R_x B$, and $A_{1(2)}=\textrm{Im}[K_{p(s)0}]L_{\textrm{Diff}}$. Here $L_{\textrm{Diff}}=\omega_pR_x^2/c$, $\tau_0$, and $U_0$ are typical diffraction length, pulse duration, and Rabi frequency of the probe pulse, respectively.

Note that for simplicity the group-velocity dispersion (i.e., the term proportional to $\partial^2 u_j/\partial \tau^2$) in Eq.~(\ref{NLSE3}) has not been neglected. Such approximation is valid for large $\tau_0$, in which case the dispersion length of the system is much larger than the diffraction and nonlinearity lengths, as shown in the following section.

\section{Trapping of weak signal pulses by soliton} {\label{Sec:4}}

\subsection{Ultraslow and matched group velocities}{\label{Sec:4a}}

Because the system under study is a lifetime broadened one, the coefficients in Eq.~(\ref{NLSE3}) are complex, which means that exact and stable nonlinear localized solutions do not exist generally. However, under the EIT condition the imaginary parts of these coefficients can be made much smaller than their real parts. Thus it is possible to obtain shape-preserving nonlinear localized solutions that can propagate to a rather long distance without significant distortion, as shown below.

Following Ref.~\cite{Sun2} we consider the solution with the form
\begin{eqnarray}\label{U}
u_j(\tau,\xi,s)=g_j(\tau,s)v_j(\tau,\xi),
\end{eqnarray}
with
$g_j(\tau,s) = \left(1/\sqrt[4]{2\pi\rho_0^2}\right) \exp [-(s-\lambda_j\tau)^2/(4\rho_0^2)]
=\left(1/\sqrt[4]{2\pi\rho_0^2}\right)\exp [-(z-V_{gj}t)^2/(4\rho_0^2L_{\rm Diff}^2)]$
($j=1,\,2$), where $\rho_0$ is a free real parameter.
After integrating over the variable $s$,  Eq.~(\ref{NLSE3}) becomes
\begin{eqnarray}\label{NLSE4}
&&\left(\frac{i}{\lambda_p}\frac{\partial}{\partial
\tau}+\frac{1}{2}\frac{\partial ^2}{\partial
\xi^2}\right)u_{1}-\frac{1}{2\sqrt{\pi}\rho_0}w_{11}|u_{1}|^2u_{1}+\mathcal{M}_1 \xi u_1=0,\\
&&\left(\frac{i}{\lambda_s}\frac{\partial}{\partial
\tau}+\frac{1}{2}\frac{\partial ^2}{\partial
\xi^2}\right)u_{2}-\frac{1}{2\sqrt{\pi}\rho_0}w_{21}|u_{1}|^2u_{2}+\mathcal{M}_2 \xi u_2=0,
\end{eqnarray}
($j=1,2$). By taking realistic  physical parameters $U_0=7.75\times10^6$ s$^{-1}$, $\tau_0=2.75\times10^{-7}$ s, and $R_x=25$ $\mu$m and the other parameters the same as those given in the above two sections, we obtain $\lambda_1\approx\lambda_2=1.0$, $w_{11}\approx w_{21}= -1.0-0.18i$, $M_1=(2.45+i0.0016)\times10^{4}$ mG$^{-1}$cm$^{-1}$, and $M_2\approx 0$ due to the selected energy-level structure (see the end  of Sec.~\ref{Sec:2}). We see that the imaginary parts of the coefficients in Eq.~(\ref{NLSE3}) are indeed much smaller than their corresponding real parts. Based on these parameters, we obtain $L_{\rm Diff}=0.49$ cm, which is approximately equal to  $L_{\rm Nonl}(\equiv1/U_0^2|W_{11}|$; typical nonlinearity length). However, the typical linear absorption length $L_{{\rm A}j}=1/{\rm Im}(K_{p(s)0})$ is around $1738.4$ cm and the typical dispersion length $L_{{\rm Disp}j}=\tau_0^2/{\rm Re}(K_{p(s)2})$ [$K_{p(s)2}\equiv \partial^2 K_{p(s)}/\partial \omega^2$], is estimated to be $3.76$ cm, both of them are much larger than $L_{\rm Diff}$ and $L_{\rm Nonl}$. Based on the results, we thus have the group-velocity dispersion ratios $d_j=L_{\rm Diff}/L_{{\rm Disp}j}\approx0.13$ and the absorption coefficients $A_j=L_{\rm Diff}/L_{{\rm A}j}\approx0.00028$, which show the significance of the various typical interaction lengths relative to the diffraction effect. Therefore, the corresponding terms in Eq.~(\ref{NLSE3}) can be indeed negligible.

With the above parameters we obtain the group velocities of the probe and signal pulses
\begin{subequations}
\begin{eqnarray}
&&\tilde{V}_{g1}={\rm Re}\left(\frac{\partial K_p}{\partial \omega}\right)^{-1}=5.978\times10^{-5}\,c,\\
&&\tilde{V}_{g2}={\rm Re}\left(\frac{\partial K_s}{\partial \omega}\right)^{-1}=5.980\times10^{-5}\,c,
\end{eqnarray}
\end{subequations}
respectively. We see the following: (i) both the group velocities are ultraslow (i.e., much smaller than $c$); (ii) both of them have nearly the same value (i.e., matched each other). Note that the ultraslow and matched group velocities are very important for long interaction time between the probe and signal pulses and hence for an efficient CPM, which is essential for realizing the trapping of weak signal pulses by soliton.

\subsection{One weak signal pulse trapped by probe soliton}{\label{Sec:4b}}

We now study the trapping of the weak signal pulse by a probe soliton without the SG magnetic field
(i.e., $B=0$ and hence $\mathcal{M}_j=0$). In this case, Eq.~(\ref{NLSE4}) is simplified as
\begin{subequations}\label{NLSE5}
\begin{eqnarray}
&& \label{NLSE51}\left(\frac{i}{\lambda_1}\frac{\partial}{\partial
\tau}+\frac{1}{2}\frac{\partial ^2}{\partial
\xi^2}\right)v_{1}-\frac{1}{2\sqrt{\pi}\rho_0}w_{11}|v_{1}|^2v_{1}=0,\\
&&\label{NLSE52}\left(\frac{i}{\lambda_2}\frac{\partial}{\partial
\tau}+\frac{1}{2}\frac{\partial ^2}{\partial
\xi^2}\right)v_{2}-\frac{1}{2\sqrt{\pi}\rho_0}w_{21}|v_{1}|^2v_{2}=0.
\end{eqnarray}
\end{subequations}
Here $v_1$ ($v_2$) is related to the probe (signal) field.

Note that (\ref{NLSE51}) is a
nonlinear Schr\"{o}dinger (NLS) equation, whose
solution is well known. The single-soliton solution reads

\begin{eqnarray}\label{soliton}
u_1(\tau,\xi)=\varsigma_1{\rm sech}({\sqrt{2q_1}}\varsigma_1\xi/ {2})\exp(i \varsigma_1^2q_1 \lambda_p\tau/{4}),
\end{eqnarray}
where $q_1=-w_{11}/(\sqrt{\pi}\rho_0)$, and $\varsigma_1$, $p$, and $n$ are free parameters.
The solution of Eq.~(\ref{NLSE52}) depends on the solution of Eq.~(\ref{NLSE51}) because $v_1$ plays a role of ``external potential" in Eq.~(\ref{NLSE52}). Thus one expects that a trapping of $v_2$ can be realized by the probe soliton given by $v_1$. Since the free parameter $p$ is trivial by the Galilean invariance of  Eq.~(\ref{NLSE5}), we discuss only the case of $p=0$ below.

Based on the form of the soliton solution (\ref{soliton}), the analytical solution for $v_2$ can be easily
obtained~\cite{MF}. We assume the solution of Eq.~(\ref{NLSE5}b) has the form
$v_2(\tau,\xi)=\varsigma_2h(\xi)\exp(i\beta \lambda_2\tau)$, where $\varsigma_2$
and $\beta$ are the amplitude and propagation constants.
Substituting this form into Eq.~(\ref{NLSE52}), we obtain
$-\frac{1}{2}\frac{\partial ^2}{\partial \xi^2}h+U(\xi)h=-\beta h$,
where $U(\xi)=-q_2\varsigma_1^2{\rm sech}^2\left(\sqrt{2q_1}\varsigma_1\xi/2\right)$ with
$q_2=-w_{21}/(2\sqrt{\pi}\rho_0)$. Here $U(\xi)$ represents the potential well, $h(\xi)$ is the eigenfunction, and $-\beta$ is the corresponding eigenvalue that represents a discrete energy level. The fundamental (even) mode solution is given by $h_0(\xi)={\rm sech}^{n_0}(a\xi)$, with $a=\sqrt{2q_1}\varsigma_1/2$, $\beta=q_1\varsigma_1^2n_0^2/4$,
and $n_0=(-q_1+\sqrt{q_1^2+16q_1q_2})/2q_1$. The first-order (odd) mode solution reads
$h_1(\xi)={\rm sech}^{n_1}(a\xi)\tanh(a\xi)$, with $\beta=q_1\varsigma_1^2n_1^2/4$ and $n_1=n_0-1$.
One can find two localized modes for this system analytically.
Note that $\varsigma_2$ can be taken as an arbitrary value as long as $\varsigma_2\ll \varsigma_1$
for either the fundamental or the first-order mode.
The parameters $q_1$ and $q_2$ which affect an eigenfunction are determined by system parameters.

%
\begin{figure}
\includegraphics[scale=0.7]{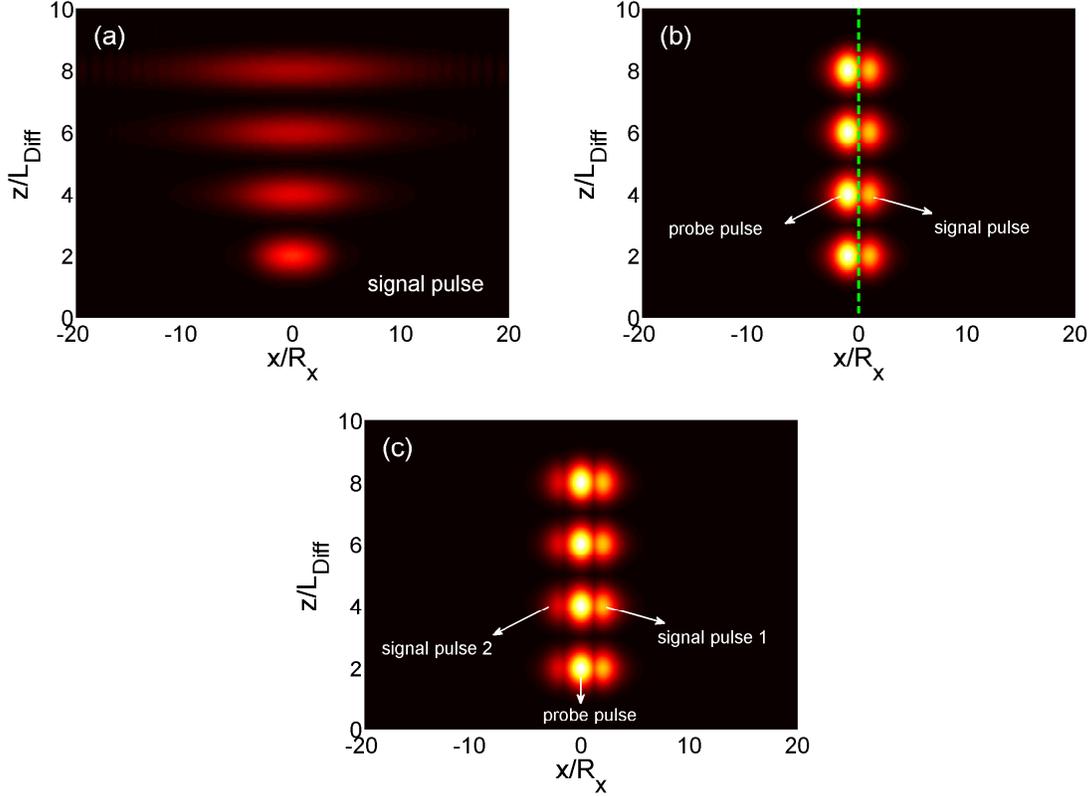}
\caption{(Color online) Trapping of weak signal pulses by probe soliton pulse. (a) The spreading of the signal pulse when the probe field is absent. (b) The signal pulse is
trapped by the probe soliton pulse. The probe and signal pulses propagate together stably along the line $x=0$. The green dashed line in the panel is the central position of the probe soliton pulse calculated theoretically. (c) Two signal pulses (i.e., the signal pulse 1 and the signal pulse 2)  trapped by the probe soliton pulse. The spots from bottom to top in all the panels of the figure represent $|\Omega_p/U_0|$ and $|\Omega_s/U_0|$ for the pulses propagating to $z=2L_{\rm Diff}$, $z=4L_{\rm Diff}$, $z=6L_{\rm Diff}$, and $z=8L_{\rm Diff}$, respectively.}\label{fig:3}
\end{figure}

For simplicity, we consider only the even mode (i.e. bright soliton) of $v_2$. Shown in Fig.~\ref{fig:3}(a) is the propagation of the signal pulse when the probe field is absent (i.e., $v_1=0$). In this case, no trapping of the signal pulse happens and the signal pulse spreads rapidly (i.e., unstable) during propagation. Figure~\ref{fig:3}(b) shows the case when the probe pulse with the form of soliton (\ref{soliton}) is present, with $\varsigma_1=1.2$. The initial condition of the signal pulse is assumed to have the
form $v_2=\varsigma_2{\rm sech}(\varsigma_1\xi)$ with $\varsigma_2=0.2$. We see that a trapping of the signal pulse occurs. In this situation, the probe soliton contributes a CPM effect (and thus a trapping potential) to the signal pulse. As a result, the spreading of the signal pulse is arrested and both the probe and the signal pulses propagate together stably along the line $x=0$ with the same velocity. The green dashed line in the panel (b) is the propagating path of the probe soliton calculated theoretically. Note that the bright light spots from the bottom in the figure to the top represent $|\Omega_p/U_0|$ (the probe field) and $|\Omega_s/U_0|$ (the signal field) for the pulses propagating to $z=2L_{\rm Diff}$, $z=4L_{\rm Diff}$, $z=6L_{\rm Diff}$, and $z=8L_{\rm Diff}$, respectively. Note that for a better visualization the intensity of the signal pulse plotted in Fig.~\ref{fig:3} has been amplified four times, and the central positions of the probe soliton and the signal pulses have been separated a small distance artificially (the same treatment is also used in Fig.~\ref{fig:4} and Fig.~\ref{fig:5}).

The trapping phenomenon predicted above can be used to design an all-optical switcher. The principle is the following. Assume the signal pulse passes through the medium with only a diffraction effect; then it detrimentally expands in the $x$ direction when the probe soliton is absent, which corresponds to the switch-off of the switcher; see Fig.~\ref{fig:3}(a). When the probe soliton is present, the signal pulse is trapped by the probe soliton and propagates along a trajectory without distortion, as shown in Fig.~\ref{fig:3}(b), which corresponds to the switch-on of the switcher.

\subsection{Multiple signal pulses trapped by the probe soliton pulse} {\label{Sec:4c}}

The above theory can be generalized to the case of multiple signal pulses
trapped by a probe soliton pulse, which can be obtained by considering a $(N+1)$-pod
system with level diagram and excitation scheme the same as Fig.~\ref{fig:1}(a), but with
one probe pulse and $N-1$ signal pulses. Similar to that done in Sec.~{\ref{Sec:4a}},
we can obtain the following coupled envelope equations
\begin{eqnarray}\label{NLSE6}
i\left(\frac{\partial}{\partial
z_2}+\frac{1}{V_{\textrm{g}j}}\frac{\partial}{\partial
t_2}\right)F_{j}+\frac{c}{2\omega_{pj}}\frac{\partial ^2}{\partial
x_1^2}F_{j}-W_{j1}|F_{1}|^2F_{j}=0
\end{eqnarray}
$(j=1,\,2,\,\ldots,\,N)$. Here, $F_j$ is the envelope of the $j$th pulse; $W_{j1}$ are coefficients of self-phase (for $j$=1) and cross-phase (for $j\neq$1) modulations, which are enhanced by the EIT effect induced by the control field. Explicit expressions of these coefficients are lengthy and omitted here. When deriving Eq.~(\ref{NLSE6}), the SG gradient magnetic field is not included. Note that due to the symmetry of the system, the group velocities $V_{\textrm{g}j}$ ($j=1,2,\cdots,N$) of all pulses are ultraslow and nearly matched.

We can study the solutions of Eq.~(\ref{NLSE6}) and discuss the
trapping phenomenon in such $(N+1)$-pod system based on the approach given in the last section.
For simplicity, here we consider the case $N=3$. A general consideration for $N>3$ can be done in a similar way. Shown in Fig.~\ref{fig:3}(c) is the result for $N=3$ by numerical simulation, where two signal pulses (i.e. the signal pulse 1 and the signal pulse 2) are trapped by a one probe soliton. We see that all the pulses propagate with the same velocity, and keep their profiles unchanged. Such phenomenon of multiple weak pulses trapping by a soliton pulse can also be employed to design all-optical switching.


\section{Trajectory control of trapped signal pulse and soliton}{\label{Sec:5}}

\subsection{SG deflection of trapped signal pulse and soliton}{\label{Sec:5a}}

We now turn to explore the possibility of trajectory control of the trapped signal pulse and soliton by means of an external field. We consider the tripod atomic system interacted by a SG gradient magnetic field (Fig.~\ref{fig:1}). For clearness, we write the two components of Eq.~(\ref{NLSE4}) explicitly:
\begin{subequations} \label{EPE}
\begin{eqnarray}
&&\label{EPE1}\left(\frac{i}{\lambda_1}\frac{\partial}{\partial \tau}+\frac{1}{2}\frac{\partial^2}{\partial \xi^2}\right)v_1-\frac{1}{2\sqrt{\pi}\rho_0}w_{11}|v_1|^2v_1+\mathcal{M}_1\xi v_1=0,\\
&&\label{EPE2}\left(\frac{i}{\lambda_2}\frac{\partial}{\partial \tau}+\frac{1}{2}\frac{\partial^2}{\partial \xi^2}\right)v_2-\frac{1}{2\sqrt{\pi}\rho_0}w_{21}|v_1|^2v_2+\mathcal{M}_2\xi v_2=0.
\end{eqnarray}
\end{subequations}
Equation~(\ref{EPE1}) is a NLS equation with an external potential proportional to $\xi$, which can be solved exactly by using the transformation~\cite{Huang1,Huang2,HHChen}
\begin{equation}
\label{TRANS}\tau^{\prime}=\tau,\,\,\,\,\,\,\,\,\,\xi^{\prime}=\xi-\mathcal{M}_1\lambda_1^2\tau^2/2,
\end{equation}
and $v_1=\phi_1 \exp{[i\mathcal{M}_1\lambda_1\tau(\xi-\mathcal{M}_1\lambda_1^2\tau^2/6)]}$.
Under such transformation, Eq.~(\ref{EPE}a) is converted to the ``free'' NLS equation:
\begin{eqnarray}\label{TE1}
&&\left(\frac{i}{\lambda_1}\frac{\partial}{\partial \tau^{\prime}}+\frac{1}{2}\frac{\partial^2}{\partial \xi^{\prime2}}\right)\phi_1-\frac{1}{2\sqrt{\pi}\rho_0}w_{11}|\phi_1|^2\phi_1=0.
\end{eqnarray}
However, under the above transformation, Eq.~(\ref{EPE}b) is transferred into another but complicated form, which is omitted here for saving space.

The single-soliton solution of $\phi_1$ is given by
\begin{eqnarray}\label{TE1S}
\phi_1(\xi^{\prime},\tau^{\prime})=\varsigma_1 {\textrm {sech}}\left[\frac{\sqrt{2q_1}}{2}\varsigma_1\left(\xi^{\prime}-\frac{p\lambda_1\tau}{2}\right)\right]
\exp\left[i\frac{p}{2}\left(\xi^\prime-\frac{p\lambda_1\tau}{2}\right)+i n \frac{\lambda_1\tau}{2}\right],
\end{eqnarray}
where $q_1=-w_{11}/(\sqrt{\pi}\rho_0)$, and $\varsigma_1$, $p$, and $n$ are constants. We see that the probe soliton (\ref{TE1S}) moves along a parabolic trajectory
because $\xi^{\prime}$ is proportional to $\tau^2$. Since the analytical solution of $v_2$ is not available, we resort to numerical calculation. We expect that $v_2$ will be trapped by the soliton $v_1$ because $v_1$ contributes a trapping potential to $v_2$.

\begin{figure}
\includegraphics[scale=0.7]{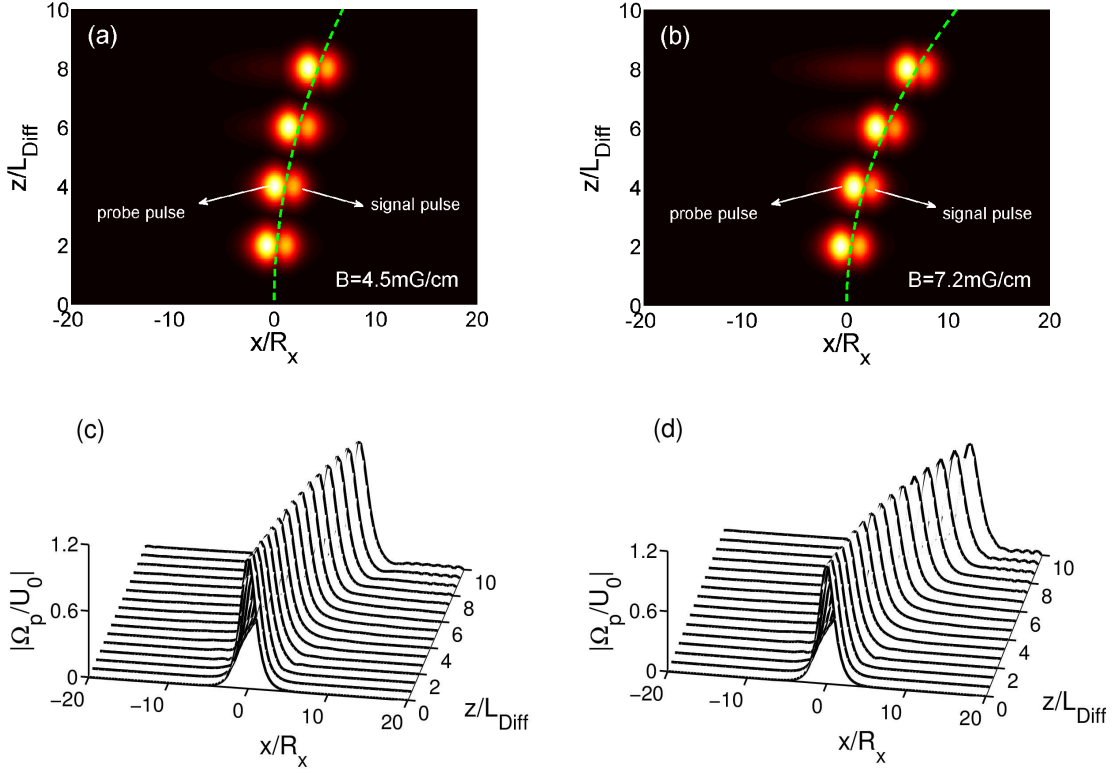}
\caption{(Color online) Stern-Gerlach deflection of the trapped signal pulse and soliton.
(a) Traveling trajectory of the probe soliton pulse and the trapped signal pulse for SG magnetic field gradient $B=4.5$ mG/cm.  (b) The same as (a) but with $B=7.2$ mG/cm. In all panels, the bright light spots from bottom to top represent $|\Omega_p/U_0|$ and $|\Omega_s/U_0|$ for the pulses propagating to $z=2L_{\rm Diff}$, $z=4L_{\rm Diff}$, $z=6L_{\rm Diff}$, and $z=8L_{\rm Diff}$, respectively; the green dashed line is the theoretical result of the trajectory of the central position of the probe soliton pulse. Panels (c) and (d) show the propagation of the probe soliton in a three-dimensional view.}\label{fig:4}
\end{figure}

Shown in Fig.~\ref{fig:4} is the numerical result of the probe and signal pulses moving in the $x$-$z$ plane
based on Eq.~(\ref{NLSE3}). Since the free parameter $p$ is trivial by Galilean invariance of the system, we take $p=0$ in the simulation. Figure~\ref{fig:4}(a) [Fig.~\ref{fig:4}(b)\,] shows the propagation of the probe soliton and the signal pulse for $B=4.5$ mG/cm ($B=7.2$ mG/cm). We see that the signal pulse is not only trapped by the probe soliton, but also undergoes a SG deflection as does the probe soliton. Both the signal pulse and the soliton propagate stably along a parabolic trajectory bent to positive $x$ direction. In the panels, the bright light spots from bottom to top represent $|\Omega_p/U_0|$ and $|\Omega_s/U_0|$ for the pulses propagating to $z=2L_{\rm Diff}$, $z=4L_{\rm Diff}$, $z=6L_{\rm Diff}$, and $z=8L_{\rm Diff}$, respectively. The green dashed line is the theoretical result of the trajectory of the probe soliton. Physically, the trapping of the signal pulse is due to the CPM by the probe soliton, whereas the SG deflection of both the signal pulse and the probe soliton is contributed by the SG gradient magnetic field. Figure~\ref{fig:4}(c) [Fig.~\ref{fig:4}(d)\,] shows the propagation of the probe soliton in a three-dimensional view corresponding to Fig.~\ref{fig:4}(a) [Fig.~\ref{fig:4}(b)\,], respectively. We see that the probe soliton radiates small-amplitude continuous waves when propagating to a large distance. Obviously, the trajectory of the trapped signal pulse and soliton can be manipulated by manipulating the SG gradient magnetic field. Such manipulation is useful for optical information processing, e.g., for the control of the behavior of all-optical switching.

\subsection{Trajectory control by using a time-dependent SG gradient magnetic field}

In the above calculation, the magnetic field $B$ is taken to be static. We now consider what will happen when a time-dependent SG gradient magnetic field is applied. In this case, the parameter $B$ in Eq.~(\ref{B}) is replaced by $B f(t)$. Then Eqs.~(\ref{EPE1}) and (\ref{EPE2})
become
\begin{subequations}\label{TNLSE}
\begin{eqnarray}
&&\label{TNLSE1}\left(\frac{i}{\lambda_1}\frac{\partial}{\partial \tau}+\frac{1}{2}\frac{\partial^2}{\partial \xi^2}\right)v_1-\frac{1}{2\sqrt{\pi}\rho_0}w_{11}|v_1|^2v_1+\mathcal{M}_1f(\tau)\xi v_1=0,\\
&&\label{TNLSE2}\left(\frac{i}{\lambda_2}\frac{\partial}{\partial \tau}+\frac{1}{2}\frac{\partial^2}{\partial \xi^2}\right)v_2-\frac{1}{2\sqrt{\pi}\rho_0}w_{21}|v_1|^2v_2+\mathcal{M}_2f(\tau)\xi v_2=0.
\end{eqnarray}
\end{subequations}
By using the transformation
\begin{subequations}
\begin{eqnarray}
&&\label{TRANS}\tau^{\prime}=\tau,\,\,\,\,\,\,\,\,\,\xi^{\prime}=\xi+\eta(\tau^{\prime}),\\
&&v_1(\xi,\tau)=\psi_1(\xi^\prime,\tau^\prime) \exp{\{i[\alpha(\tau^{\prime})\xi^{\prime}-\beta(\tau^{\prime})]\}},
\end{eqnarray}
\end{subequations}
where $\alpha=\lambda_1\mathcal{M}_1 \int f(\tau^\prime)d \tau^\prime$, $\eta=-\lambda_1
\int \alpha d \tau^\prime$, and $\beta=\lambda_1 \int [\mathcal{M}_1f(\tau^\prime)
\eta-\alpha^2/2]d \tau^\prime$, Eq.~(\ref{TNLSE1}) becomes the NLS equation
without external potential,
\begin{eqnarray}
&&\left(\frac{i}{\lambda_1}\frac{\partial}{\partial \tau^{\prime}}+\frac{1}{2}\frac{\partial^2}{\partial \xi^{\prime2}}\right)\psi_1-\frac{1}{2\sqrt{\pi}\rho_0}w_{11}|\psi_1|^2\psi_1=0,
\end{eqnarray}
with the single soliton solution given by
\begin{eqnarray}\label{SOL}
\psi_1(\xi^{\prime},\tau^{\prime})=\varsigma_1 {\textrm {sech}}\left[\frac{\sqrt{2q_1}}{2}\varsigma_1\left(\xi^{\prime}-\frac{p\lambda_1\tau}{2}\right)\right]
\exp\left[i\frac{p}{2}\left(\xi^\prime-\frac{p\lambda_1\tau}{2}\right)+in \frac{\lambda_1\tau}{2}\right],
\end{eqnarray}
where $q_1=-w_{11}/(\sqrt{\pi}\rho_0)$, and $\varsigma_1$, $p$, and $n$ are free constants.
We see that the probe pulse is still a bright soliton with the trajectory depending on the form of $f(\tau)$.
To show the evolution of the corresponding trapping phenomenon, we consider the following two simple cases:

(i) The sinusoidally oscillating external force with the form
\begin{equation}\label{OSC}
f(\tau)=A\cos(C\tau)
\end{equation}
is added to the system, where $A$ and $C$ are free parameters.
In this situation, the solution (\ref{SOL}) with $p=0$ reads
\begin{eqnarray}\label{SOL1}
\psi_1(\xi,\tau)=\varsigma_1 {\textrm {sech}}\left[\frac{\sqrt{2q}}{2}\varsigma_1\left(\xi+\frac{\lambda_1^2\mathcal{M}_1A}{C^2}\cos(C\tau)
+\eta_0\right)\right]
\exp\left(in \frac{\lambda_1\tau}{2}\right),
\end{eqnarray}
where $\eta_0=-\lambda_1^2\mathcal{M}_1A/C^2$.  Shown in Fig.~\ref{fig:5}(a)
%
\begin{figure}
\includegraphics[scale=0.7]{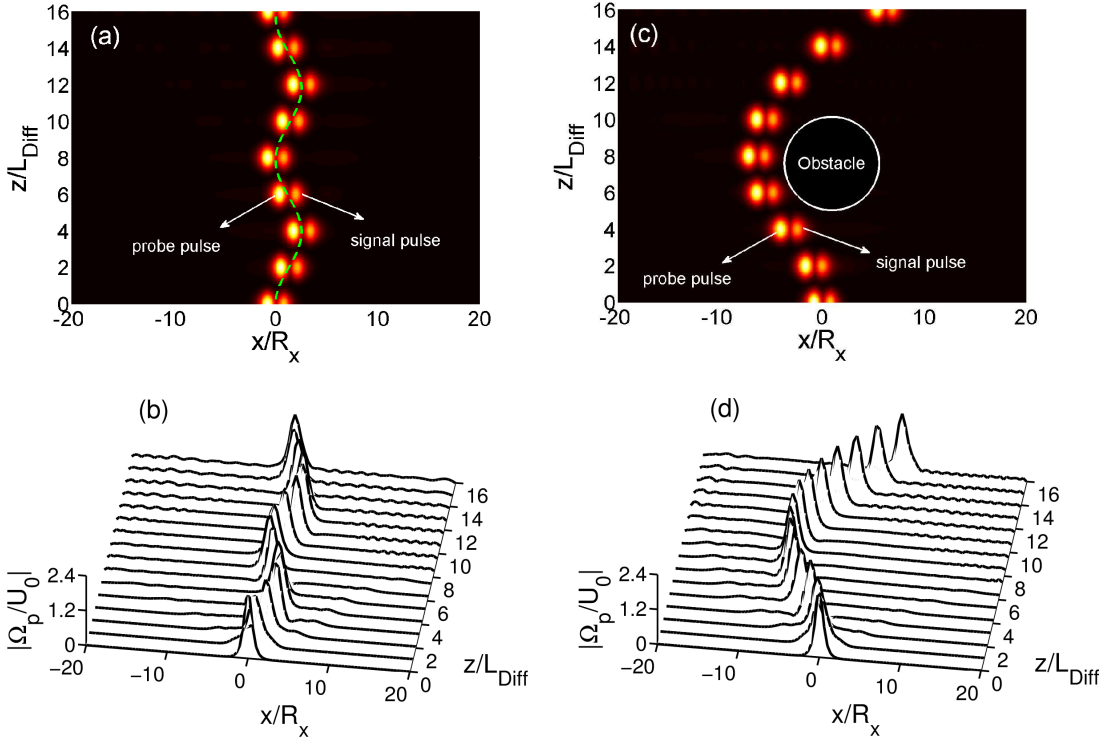}
\caption{(Color online) Trajectory control of the trapped signal pulse and soliton via time-dependent SG magnetic field. (a) The trajectory of the trapped signal pulse and soliton oscillates sinusoidally in the $x$-$z$ plane. The green dashed line is the theoretical result of the central position of the probe soliton pulse. Corresponding three-dimensional diagram is given in panel (b). (c) The trapped signal pulse and soliton bypasses an obstacle, where the black circle represents the obstacle. Corresponding three-dimensional diagram is given in panel (d). In the panels (a) and (c), the bright light spots from bottom to top indicate $|\Omega_p/U_0|$ and $|\Omega_s/U_0|$ when the pulses propagate to $z=0,2L_{\rm Diff}, 4L_{\rm Diff}, 6L_{\rm Diff}, 8L_{\rm Diff}, 10L_{\rm Diff}, 12L_{\rm Diff}, 14L_{\rm Diff}$, and $16L_{\rm Diff}$, respectively.
}\label{fig:5}
\end{figure}
%
is the numerical result of the trajectories of the probe and signal pulses in the $x$-$z$ plane. We see that due to the CPM effect the probe soliton can trap the signal pulse and both of their trajectories oscillate sinusoidally.
The bright light spots from bottom to top indicate $|\Omega_p/U_0|$ and $|\Omega_s/U_0|$ for the pulses propagating to $z=0,2L_{\rm Diff}, 4L_{\rm Diff}, 6L_{\rm Diff}, 8L_{\rm Diff}, 10L_{\rm Diff}, 12L_{\rm Diff}, 14L_{\rm Diff}$, and $16L_{\rm Diff}$, respectively. The green dashed line is the theoretical result of the central position of the probe soliton, which reads $z=(V_g\tau_0/C)\arccos[-(x+\eta_0R_x)C^2/(A\mathcal{M}_1R_x\lambda_1^2)]$. In the simulation, we have taken $B=13.5$ mG/cm, $A=2$, $C=0.8$, $\varsigma_1=2.4$ (the amplitude of the probe soliton), and $\varsigma_2=0.4$ (the amplitude of signal pulse). Figure~\ref{fig:5}(b) gives the corresponding three-dimensional diagram, which shows that there are some small radiations appearing on the tail of the soliton when the probe soliton propagates to a large distance.

(ii) The tangent-type external force with the form
\begin{equation}\label{TDM}
f(\tau)=\tanh[2(\tau-\tau_1)]
\end{equation}
is applied to the system, where $\tau_1$ is the time at which the SG gradient magnetic field changes its sign.

Equations (\ref{TNLSE1}) and (\ref{TNLSE2}) are solved numerically with $\tau_1=4$, with the result
shown in Fig.~\ref{fig:5}(c). We see that the probe soliton can still trap the signal pulse and, interestingly,  both of them travel together along a $\Lambda$-shaped path and hence can bypass an obstacle (indicated by the black circle in the figure). Illustrated in Fig.~\ref{fig:5}(d) is the propagation of the probe soliton bypassing an obstacle in a three-dimensional view, which shows that there are also some small radiations during the propagation of the soliton.

The above trajectory control technique of the trapping phenomenon can also be generalized to the $(N+1)$-pod system, where the probe soliton  can trap $N-1$ signal pulses (Sec.~\ref{Sec:4c}). In a similar way, we can use time-dependent SG gradient magnetic fields to guide the probe soliton and the $N-1$ trapped signal pulses, and make all of them oscillate or bypass an obstacle together.

Finally, we make an estimation on $P_{\rm probe}$ (the power of the probe
soliton) and $P_{\rm signal}$ (the power of the signal pulse).
Using Poynting's vector~\cite{HDP}, it is easy to obtain
\begin{subequations}
\begin{eqnarray}
&&P_{\rm probe}\approx1.6\times10^{-8} W,\\
&&P_{\rm signal}\approx4.9\times10^{-10} W,
\end{eqnarray}
\end{subequations}
both of which are at ultraweak light level. This is very different from conventional optical media, such as glass-based optical fibers, where picosecond or femtosecond laser pulses are usually needed to reach a high power to bring out a sufficiently nonlinear effect needed for the formation of an optical soliton~\cite{NG,DGC,GS1,GS2,MF}.

\section{Summary}\label{Sec:6}

In this article, we have proposed a mechanism for trapping weak signal pulses
by using a probe soliton and realizing its trajectory control via EIT. By means of the EIT, not only the optical absorption can be largely suppressed but also the enhanced Kerr effect can be obtained. We have shown that several weak signal pulses can be easily trapped by a probe soliton and stably cotravel with ultraslow  propagating velocity. Furthermore, we have demonstrated that the trajectories of the probe soliton and the trapped signal pulses can be steered by using a SG gradient magnetic field.  As a result, the probe soliton and the trapped signal pulses display a SG deflection and both of them can bypass an obstacle together. The results predicted here may be useful for light information processing, such as for the design of all-optical switching at very low light level.


\begin{acknowledgments}

This work was supported by NSF-China under Grants No. 10874043 and and No. 11174080.

\end{acknowledgments}


\appendix

\section{EQUATIONS OF MOTION FOR THE DENSITY-MATRIX ELEMENTS}\label{AppendixA}

Explicit expressions of the equations of motion for the density matrix elements $\sigma_{jl}$
are
\begin{subequations} \label{dme}
\begin{align}
&i\frac{\partial}{\partial
t}\sigma_{11}-i\Gamma_{13}\sigma_{33}-i\Gamma_{14}\sigma_{44}+\Omega_{p}^{\ast}\sigma_{41}
-\Omega_{p}\sigma_{41}^{\ast}=0,\\
&i\frac{\partial}{\partial
t}\sigma_{22}-i\Gamma_{23}\sigma_{33}-i\Gamma_{24}\sigma_{44}+\Omega_{s}^{\ast}\sigma_{42}
-\Omega_{s}\sigma_{42}^{\ast}=0,\\
&i\left(\frac{\partial}{\partial
t}+\Gamma_{3}\right)\sigma_{33}-i\Gamma_{34}\sigma_{44}+\Omega_{c}^{\ast}\sigma_{43}
-\Omega_{c}\sigma_{43}^{\ast}=0,\\
&i\left(\frac{\partial}{\partial
t}+\Gamma_{4}\right)\sigma_{44}+\Omega_{p}\sigma_{41}^{\ast}+\Omega_{s}\sigma_{42}^{\ast}
+\Omega_{c}\sigma_{43}^{\ast}-\Omega_{p}^{\ast}\sigma_{41}-\Omega_{s}^{\ast}\sigma_{42}
-\Omega_{c}^{\ast}\sigma_{43}=0,\\
&\left(i\frac{\partial}{\partial
t}+d_{21}\right)\sigma_{21}+\Omega_{s}^{\ast}\sigma_{41}-\Omega_{p}\sigma_{42}^{\ast}=0,\\
&\left(i\frac{\partial}{\partial
t}+d_{31}\right)\sigma_{31}+\Omega_{c}^{\ast}\sigma_{41}-\Omega_{p}\sigma_{43}^{\ast}=0,\\
&\left(i\frac{\partial}{\partial
t}+d_{32}\right)\sigma_{32}+\Omega_{c}^{\ast}\sigma_{42}-\Omega_{s}\sigma_{43}^{\ast}=0,\\
&\left(i\frac{\partial}{\partial
t}+d_{41}\right)\sigma_{41}+\Omega_{p}(\sigma_{11}-\sigma_{44})+\Omega_{s}\sigma_{21}
+\Omega_{c}\sigma_{31}=0,\\
&\left(i\frac{\partial}{\partial
t}+d_{42}\right)\sigma_{42}+\Omega_{s}(\sigma_{22}-\sigma_{44})+\Omega_{p}\sigma_{21}^{\ast}
+\Omega_{c}\sigma_{32}=0,\\
&\left(i\frac{\partial}{\partial
t}+d_{43}\right)\sigma_{43}+\Omega_{c}(\sigma_{33}-\sigma_{44})+\Omega_{p}\sigma_{31}^{\ast}
+\Omega_{s}\sigma_{32}^{\ast}=0,
\end{align}
\end{subequations}
where $d_{jl}=\Delta_{j}-\Delta_{l}+i\gamma_{jl}$, $\Delta_2=\delta_2-\mu_{21}Bx$, $\Delta_3=\delta_3-\mu_{31}Bx$, and $\Delta_4=\delta_4-\mu_{41}Bx$, with $\mu_{jl}=\mu_B(g_F^jm_F^j-g_F^lm_F^l)$. Here we have defined $\delta_2=\omega_{p}-\omega_{s}-\omega_{21}$,
$\delta_3=\omega_{p}-\omega_{c}-\omega_{31}$, and
$\delta_4=\omega_{p}-\omega_{41}$, with
$\omega_{jl}=(E_j-E_l)/\hbar$  and $E_j$ the eigenenergy of
the state $|j\rangle$. Dephasing rates are defined as
$\gamma_{jl}=(\Gamma_j+\Gamma_l)/2+\gamma_{jl}^{\rm col}$, with
$\Gamma_j=\sum_{E_i<E_j}\Gamma_{ij}$  denoting the
spontaneous emission rate from the state $|j\rangle$ to all lower-energy
states $|i\rangle$ and $\gamma_{jl}^{\rm col}$ denoting the dephasing
rate reflecting the loss of phase coherence between $|j\rangle$ and
$|l\rangle$, as might occur with elastic collisions.

\section{EXPRESSIONS OF THE SOLUTIONS IN EACH-ORDER APPROXIMATIONS}\label{AppendixB}

(i) \textit{First-order approximation}:
\begin{subequations}
\begin{eqnarray}
&&\Omega_{p}^{(1)}=F_{1}e^{i\theta_p},\\
&&\sigma_{31}^{(1)}=-\frac{\Omega_c^{\ast}\sigma_{11}^{(0)}}{D_1}F_{1}e^{i\theta_p},\\
&&\sigma_{41}^{(1)}=\frac{(\omega+d_{31}^{(0)})\sigma_{11}^{(0)}}{D_1}F_{1}e^{i\theta_p},
\end{eqnarray}
\end{subequations}
where $D_1=|\Omega_c|^2-(\omega+d_{31}^{(0)})(\omega+d_{41}^{(0)})$.

(ii) \textit{Second-order approximation}:
\begin{subequations}
\begin{eqnarray}
&&\Omega_{s}^{(2)}=F_{2}e^{i\theta_s},\\
&&\sigma_{32}^{(2)}=-\frac{\Omega_c^{\ast}\sigma_{22}^{(0)}}{D_2}F_{2}e^{i\theta_s},\\
&&\sigma_{42}^{(2)}=\frac{(\omega+d_{32}^{(0)})\sigma_{22}^{(0)}}{D_2}F_{2}e^{i\theta_s},\\
&&\sigma_{jj}^{(2)}=a_{jj}^{(2)}|F_1|^2e^{-2\bar{a}_1z_2}\,\,\,\,\,\, (j=1--4),\\
&&\sigma_{43}^{(2)}=a_{43}^{(2)}|F_1|^2e^{-2\bar{a}_1z_2},
\end{eqnarray}
\end{subequations}
with $D_2=|\Omega_c|^2-(\omega+d_{32}^{(0)})(\omega+d_{42}^{(0)})$  and
\begin{subequations}
\begin{eqnarray}
&&a_{11}^{(2)}=a_{22}^{(2)}=\frac{(\Gamma_{24}-\Gamma_{23})\sigma_{11}^{(0)}}{i2(\Gamma_{13}\Gamma_{24}-\Gamma_{23}\Gamma_{14})}\left(\frac{\omega+d_{31}^{\ast(0)}}{D_1^{\ast}}-\frac{\omega+d_{31}^{(0)}}{D_1}\right),\\
&&a_{33}^{(2)}=-\frac{\Gamma_{24}\sigma_{11}^{(0)}}{i(\Gamma_{13}\Gamma_{24}-\Gamma_{23}\Gamma_{14})}\left(\frac{\omega+d_{31}^{\ast(0)}}{D_1^{\ast}}-\frac{\omega+d_{31}^{(0)}}{D_1}\right),\\
&&a_{44}^{(2)}=\frac{\Gamma_{23}\sigma_{11}^{(0)}}{i(\Gamma_{13}\Gamma_{24}-\Gamma_{23}\Gamma_{14})}\left(\frac{\omega+d_{31}^{\ast(0)}}{D_1^{\ast}}-\frac{\omega+d_{31}^{(0)}}{D_1}\right),\\
&&a_{43}^{(2)}=\frac{\Omega_c\sigma_{11}^{(0)}}{\omega+d_{43}^{(0)}}\left[\frac{1}{D_1^{\ast}}+\frac{\Gamma_{24}+\Gamma_{23}}{i(\Gamma_{13}\Gamma_{24}-\Gamma_{23}\Gamma_{14})}\left(\frac{\omega+d_{31}^{\ast(0)}}{D_1^{\ast}}-\frac{\omega+d_{31}^{(0)}}{D_1}\right)\right].
\end{eqnarray}
\end{subequations}

(iii) \textit{Third-order approximation}:
The solvability condition of $F_1$ requires
\begin{eqnarray}
i\left(\frac{\partial}{\partial
z_2}+\frac{1}{V_{\textrm{g}1}}\frac{\partial}{\partial
t_2}\right)F_{1}+\frac{c}{2\omega_{p}}\frac{\partial ^2}{\partial
x_1^2}F_{1}-W_{11}|F_{1}|^2F_{1}e^{-2\bar{a}_1z_2}+M_1 Bx_1F_1=0,
\end{eqnarray}
with
\begin{eqnarray}\label{W11}
&&W_{11}=-\kappa_{14}\frac{\Omega_ca_{43}^{\ast(2)}+(\omega+d_{31})
(a_{11}^{(2)}-a_{44}^{(2)})}{|\Omega_c|^2-(\omega+d_{31}^{(0)})(\omega+d_{41}^{(0)})},\\
&&M_1=-\kappa_{14}\frac{(\omega+d_{31}^{(0)})^2\mu_{41}+|\Omega_c|^2\mu_{31}}{2D_1^2}.
\end{eqnarray}
The third-order approximation solution reads
\begin{eqnarray}
&&\sigma_{21}^{(3)}=a_{21}^{(3)}F_1F_2^{\ast}e^{i(\theta_p-\theta_s)},
\end{eqnarray}
with
\begin{eqnarray}
&&a_{21}^{(3)}=\frac{1}{\omega+d_{21}^{(0)}}\left(\frac{\omega+d_{32}^{\ast(0)}}{D_2^{\ast}}\sigma_{22}^{(0)}
-\frac{\omega+d_{31}^{(0)}}{D_1}\sigma_{11}^{(0)}\right).
\end{eqnarray}

(iv) \textit{Fourth-order approximation}:
The solvability condition for $F_2$ requires
\begin{eqnarray}
i\left(\frac{\partial}{\partial
z_2}+\frac{1}{V_{\textrm{g}2}}\frac{\partial}{\partial
t_2}\right)F_{2}+\frac{c}{2\omega_{s}}\frac{\partial ^2}{\partial
x_1^2}F_{2}-W_{21}|F_{1}|^2F_{2}e^{-2\bar{a}_1z_2}+M_2 Bx_1F_2=0,
\end{eqnarray}
with
\begin{eqnarray}\label{W21}
&&W_{21}=-\kappa_{24}\frac{\Omega_ca_{43}^{\ast(2)}+(\omega+d_{32})(a_{22}^{(2)}-a_{44}^{(2)}
+a_{21}^{\ast(3)})}{|\Omega_c|^2-(\omega+d_{32}^{(0)})(\omega+d_{42}^{(0)})},\\
&&M_2=-\kappa_{24}\frac{(\omega+d_{32}^{(0)})^2\mu_{42}+|\Omega_c|^2\mu_{32}}{2D_2^2}.
\end{eqnarray}
%



\begin{thebibliography}{}
\bibitem{note}Experimentally, for the convenience of detection the probe and the signal fields are
usually arranged to have very small angles with respect to the propagating direction (i.e., the $z$ axis).
\bibitem{note1} The frequency and wave number of the probe field are given by $\omega_{p}+\omega$ and $k_{p}+K_p(\omega)$,
respectively. Thus $\omega=0$ corresponds to the center frequency of the probe field.
\end{thebibliography}
\end{document}